\documentclass[11pt,a4paper]{article}

\usepackage[utf8]{inputenc}
\usepackage[T1]{fontenc}
\usepackage{lmodern}
\usepackage[margin=2.5cm]{geometry}
\usepackage{amsmath,amssymb}
\usepackage{booktabs}
\usepackage{graphicx}
\usepackage{hyperref}
\usepackage{listings}
\usepackage{xcolor}
\usepackage{float}
\usepackage{caption}
\usepackage{subcaption}
\usepackage{authblk}
\usepackage{multirow}
\usepackage{url}
\usepackage{tabularx}

\hypersetup{
    colorlinks=true,
    linkcolor=blue!70!black,
    citecolor=blue!70!black,
    urlcolor=blue!70!black,
}

\lstdefinelanguage{lapis}{
    morekeywords={meta,types,ops,webhooks,errors,limits,flows,GET,POST,PUT,PATCH,DELETE},
    sensitive=true,
    morecomment=[l]{\#},
    morestring=[b]",
}

\lstdefinelanguage{yaml-openapi}{
    morekeywords={openapi,info,title,version,description,servers,url,paths,get,post,put,patch,delete,parameters,name,in,schema,type,responses,components,schemas,properties,required,items,requestBody,content,operationId,summary,security},
    sensitive=true,
    morecomment=[l]{\#},
    morestring=[b]",
}

\title{\textbf{LAPIS: Lightweight API Specification for Intelligent Systems}\\[0.3em]
\large A Token-Efficient API Description Format for LLM Context Windows}

\author[1]{Daniel Garc\'ia Garc\'ia}
\affil[1]{Independent Researcher, Spain}
\affil[ ]{\texttt{cr0hn@cr0hn.com} \quad \url{https://cr0hn.com/en/}}

\date{February 2026}

\begin{document}

\maketitle

\begin{abstract}
Large Language Models (LLMs) increasingly serve as consumers of API specifications, whether for code generation, autonomous agent interaction, or API-assisted reasoning. The de facto standard for API description, OpenAPI, was designed for documentation tools and code generators, resulting in substantial token overhead when used as LLM context. We present LAPIS (Lightweight API Specification for Intelligent Systems), a domain-specific format optimized for LLM consumption that preserves the semantic information necessary for API reasoning while minimizing token usage. Through empirical evaluation against five real-world production API specifications---including GitHub (1,080 endpoints), Twilio (197 endpoints), DigitalOcean (545 endpoints), Petstore, and HTTPBin---we demonstrate an average token reduction of 85.5\% compared to OpenAPI YAML and 88.6\% compared to OpenAPI JSON, measured with the \texttt{cl100k\_base} tokenizer. LAPIS introduces domain-specific structural innovations including centralized error definitions, webhook trigger conditions, structured rate limit descriptions, and operation flow declarations---information that OpenAPI either duplicates redundantly or cannot represent at all. The format is fully convertible from OpenAPI 3.x via an automated converter, requires no special parser for LLM consumption, and is released as an open specification under CC BY 4.0.
\end{abstract}

\noindent\textbf{Keywords:} API specification, Large Language Models, token optimization, context window, OpenAPI

\vspace{1em}

\section{Introduction}

The emergence of LLM-powered agents, AI coding assistants, and autonomous API consumers has created a new class of API specification consumer: the language model. Tools such as GitHub Copilot, Cursor, and custom agent frameworks routinely ingest API specifications to understand available endpoints, data schemas, authentication requirements, and operational constraints. The specification serves as the model's primary source of truth about the API's capabilities and contract.

The dominant format for API description, OpenAPI Specification (formerly Swagger)~\cite{openapi}, was designed in 2010--2015 for a different set of consumers: documentation renderers (Swagger UI), SDK generators, and API testing tools. These consumers benefit from explicit JSON Schema typing, exhaustive response enumeration, rich metadata (\texttt{contact}, \texttt{license}, \texttt{termsOfService}), and detailed examples. An LLM consumer, however, needs none of these. What an LLM requires is a concise, unambiguous description of what the API does, what inputs it accepts, and what outputs it returns.

The mismatch between OpenAPI's design goals and LLM consumption patterns produces measurable waste:

\begin{enumerate}
    \item \textbf{Structural overhead.} JSON Schema's nested \texttt{type}/\texttt{properties}/\texttt{required} structure produces significant token overhead for type definitions that a human would express in a single line.
    \item \textbf{Error duplication.} OpenAPI requires error responses to be defined per-operation. A 401 Unauthorized response with identical schema is duplicated in every authenticated endpoint. In the GitHub REST API specification, error responses are defined 1,594 times across 1,080 operations---an average of 1.5 error definitions per operation, with the 404 code alone appearing 531 times.
    \item \textbf{Irrelevant metadata.} Fields such as \texttt{info.contact}, \texttt{info.license}, \texttt{externalDocs}, \texttt{tags} definitions, \texttt{x-*} extensions, and multiple server definitions serve documentation and governance purposes but provide no value for LLM reasoning.
    \item \textbf{Missing operational context.} Conversely, information that would help an LLM reason about API usage---such as rate limits, webhook trigger conditions, and operation sequencing---is either absent from OpenAPI or relegated to free-text descriptions.
\end{enumerate}

This paper presents LAPIS (Lightweight API Specification for Intelligent Systems), a domain-specific format that addresses these inefficiencies. LAPIS is not a replacement for OpenAPI; it is a \emph{conversion target}. Organizations maintain their OpenAPI specification as the source of truth for documentation and tooling, and convert to LAPIS when the consumer is an LLM.

Our contributions are:

\begin{enumerate}
    \item A formal specification of the LAPIS format (v0.1.0), including syntax, grammar (EBNF), and type system (Section~\ref{sec:spec}).
    \item Deterministic conversion rules from OpenAPI 3.x to LAPIS (Section~\ref{sec:conversion}).
    \item Empirical evaluation on five production API specifications demonstrating an average 85.5\% token reduction versus OpenAPI YAML (Section~\ref{sec:evaluation}).
    \item Analysis of structural waste in real-world OpenAPI specifications (Section~\ref{sec:waste}).
\end{enumerate}

\section{Related Work}
\label{sec:related}

\subsection{API Description Languages}

The OpenAPI Specification (OAS)~\cite{openapi}, maintained by the OpenAPI Initiative under the Linux Foundation, is the dominant standard for REST API description. OAS 3.1 aligns with JSON Schema Draft 2020-12, providing rich type expressiveness at the cost of verbosity. Alternative API description languages include RAML~\cite{raml} and API Blueprint~\cite{apiblueprint}, which offer different syntactic approaches but share the same design goals: human documentation and machine code generation.

\subsection{LLM Function Calling Schemas}

Major LLM providers have independently converged on a compact function-calling schema: a function name, natural-language description, and JSON Schema parameters object. This schema is used by OpenAI~\cite{openai_fc}, Anthropic~\cite{anthropic_fc}, and Google~\cite{google_fc} for tool use. While token-efficient for individual functions, this format lacks support for type reuse, operation relationships, error semantics, and API-level metadata. It is a \emph{protocol-level} construct, not a \emph{specification format}.

\subsection{Model Context Protocol (MCP)}

The Model Context Protocol~\cite{mcp}, introduced by Anthropic, defines a transport and discovery protocol for LLM tool integration. MCP operates at the execution layer---managing tool registration, invocation, and response handling---rather than at the specification layer. LAPIS and MCP are complementary: MCP defines how an LLM calls a tool; LAPIS defines how an LLM understands an API's capabilities.

\subsection{Token-Optimized Data Formats}

TOON (Token-Oriented Object Notation)~\cite{toon} is a generic data serialization format that reduces token usage by 30--60\% compared to JSON through tabular representation of uniform arrays and YAML-like indentation. TOON operates at the data encoding layer: it can serialize \emph{any} JSON-compatible data more compactly. Applied to an OpenAPI specification, TOON would reduce the encoding overhead of JSON/YAML but would not address the structural redundancies inherent in OpenAPI's design (error duplication, schema nesting, irrelevant metadata). LAPIS operates at the specification structure layer, achieving greater reduction through domain-specific restructuring.

\subsection{OpenAPI Minification}

Tools such as LLM-OpenAPI-minifier~\cite{llm_minifier} strip unnecessary elements from OpenAPI specifications (examples, descriptions, extensions) while preserving the OpenAPI format. This approach yields moderate savings (approximately 40\% token reduction) but cannot address structural redundancies because the output must remain valid OpenAPI.

\subsection{Documentation Formats}

The \texttt{llms.txt} proposal~\cite{llmstxt} defines a convention for publishing LLM-readable documentation at well-known URLs. It provides navigational structure (links to documentation pages) rather than executable API definitions and is therefore orthogonal to LAPIS.

\section{Design Principles}
\label{sec:design}

LAPIS was designed around five principles derived from the specific requirements of LLM consumers:

\paragraph{Token minimality.} Every syntactic element must justify its token cost. The target is 70--80\% fewer tokens than equivalent OpenAPI YAML. This is not merely a compression goal; it directly affects cost (tokens are priced) and capability (context windows are finite).

\paragraph{LLM-native syntax.} The format uses natural-language-adjacent syntax---function signatures, indentation-based grouping, inline type expressions---because LLMs process these patterns more naturally than deeply nested JSON/YAML structures. The syntax resembles how a human developer would describe an API on a whiteboard.

\paragraph{Lossless conversion from OpenAPI.} The transformation from OpenAPI 3.x to LAPIS must be fully automatable with deterministic rules. No manual intervention or semantic interpretation should be required.

\paragraph{Human readability.} A developer should be able to read and understand a LAPIS document without consulting documentation. The format is self-describing.

\paragraph{Semantic completeness for reasoning.} All information that an LLM needs to reason about API usage must be preserved. Information needed only by code generators (e.g., exact JSON Schema validation keywords, XML serialization hints) is deliberately excluded.

\subsection{Non-Goals}

LAPIS explicitly does not aim to:

\begin{itemize}
    \item Replace OpenAPI as a source of truth for documentation or code generation.
    \item Define transport, execution, or discovery protocols (this is MCP's domain).
    \item Support bidirectional conversion (reconstructing a complete OpenAPI spec from LAPIS).
    \item Serve as a general-purpose data serialization format (this is TOON's domain).
\end{itemize}

\section{Specification}
\label{sec:spec}

A LAPIS document consists of up to seven sections in fixed order. Only \texttt{[meta]} and \texttt{[ops]} are required.

\begin{table}[H]
\centering
\caption{LAPIS document sections.}
\label{tab:sections}
\begin{tabular}{@{}llp{7.5cm}@{}}
\toprule
\textbf{Section} & \textbf{Required} & \textbf{Purpose} \\
\midrule
\texttt{[meta]}     & Yes & API name, base URL, version, authentication \\
\texttt{[types]}    & No  & Reusable type definitions (objects, enums) \\
\texttt{[ops]}      & Yes & API operations (endpoints) \\
\texttt{[webhooks]} & No  & Push events with trigger conditions \\
\texttt{[errors]}   & No  & Error definitions, declared once globally \\
\texttt{[limits]}   & No  & Rate limits, quotas, size constraints \\
\texttt{[flows]}    & No  & Multi-step operation sequences \\
\bottomrule
\end{tabular}
\end{table}

\subsection{Meta Section}

The \texttt{[meta]} section contains API-level metadata in key-value format:

\begin{lstlisting}[language=lapis]
[meta]
api: Invoice Service
base: https://api.example.com/v2
version: 2.1.0
desc: Invoice, customer, and payment management
auth: bearer header:Authorization
\end{lstlisting}

Authentication is described with a compact syntax: \texttt{<scheme> [location:name]}. Supported schemes include \texttt{bearer}, \texttt{apikey}, \texttt{basic}, \texttt{oauth2}, and \texttt{none}.

\subsection{Type System}

LAPIS supports eight scalar types (\texttt{str}, \texttt{int}, \texttt{float}, \texttt{bool}, \texttt{date}, \texttt{datetime}, \texttt{file}, \texttt{any}) with modifiers for arrays (\texttt{[T]}), maps (\texttt{\{str:T\}}), optionality (\texttt{?}), and default values (\texttt{= value}).

Object types are defined with indentation:

\begin{lstlisting}[language=lapis]
Invoice:
  id: str
  customer_id: str
  status: InvoiceStatus
  lines: [InvoiceLine]
  total: float
  metadata?: {str:any} @since:2.1
\end{lstlisting}

Enumerations use pipe-delimited syntax on a single line:

\begin{lstlisting}[language=lapis]
InvoiceStatus: draft | sent | paid | overdue
\end{lstlisting}

Field-level annotations support versioning (\texttt{@since:X.Y}) and deprecation (\texttt{@deprecated "reason"}).

\subsection{Operations}

Operations use a signature-based syntax with directional markers for inputs (\texttt{>}) and outputs (\texttt{<}):

\begin{lstlisting}[language=lapis]
create_invoice POST /invoices
  Creates an invoice for a customer.
  > customer_id: str
  > lines: [InvoiceLine]
  > billing_address?: Address
  < Invoice
\end{lstlisting}

Parameter location follows default rules: path parameters are inferred from \texttt{\{param\}} in the route; GET/DELETE default to \texttt{@query}; POST/PUT/PATCH default to \texttt{@body}. Explicit \texttt{@location} annotations override defaults. Operation modifiers (\texttt{+paginated}, \texttt{+idempotent}, \texttt{+stream}, \texttt{+deprecated}) are appended to the signature line.

\subsection{Webhooks}

The \texttt{[webhooks]} section describes push events with an arrow syntax and trigger conditions:

\begin{lstlisting}[language=lapis]
invoice_paid -> POST /webhooks/invoice-paid
  ! When invoice.status changes to "paid".
  < event_id: str @header:X-Event-ID
  < invoice_id: str
  < amount: float
\end{lstlisting}

The \texttt{!} prefix declares trigger conditions---semantic information that OpenAPI cannot represent structurally.

\subsection{Centralized Errors}

Errors are defined once in the \texttt{[errors]} section with optional operation binding:

\begin{lstlisting}[language=lapis]
[errors]
# Base structure: ApiError
401 unauthorized
  Token is missing, expired, or invalid.
404 not_found
  Resource not found.
409 duplicate_customer @ops:create_customer
  A customer with this email already exists.
  ~ existing_customer_id: str
\end{lstlisting}

The \texttt{\~{}} prefix denotes error-specific response fields. Errors without \texttt{@ops} binding are global; errors with \texttt{@ops} are scoped to specific operations.

\subsection{Rate Limits}

The \texttt{[limits]} section provides structured rate limit information:

\begin{lstlisting}[language=lapis]
[limits]
on_exceed: 429 retry_after
plan: free
  rate: 60/m @key
  quota: 1000/mo @key "monthly requests"
plan: pro
  rate: 600/m @key
\end{lstlisting}

This structured representation enables LLMs to reason about throttling, retry strategies, and tier selection---information typically buried in free-text documentation.

\subsection{Operation Flows}

The \texttt{[flows]} section declares multi-step operation sequences:

\begin{lstlisting}[language=lapis]
invoice_lifecycle "Invoice lifecycle"
  create_invoice -> update_invoice* -> send_invoice
    -> ...(awaiting payment) -> invoice_paid | invoice_overdue
  ? invoice_paid: payment received before due date
  ? invoice_overdue: due date passes without payment
\end{lstlisting}

Flow syntax supports sequencing (\texttt{->}), repetition (\texttt{*}), branching (\texttt{\textbar}), conditional annotations (\texttt{?}), and waiting (\texttt{...()}). This provides LLMs with the operational context needed to plan multi-step API interactions.

\subsection{Formal Grammar}

The complete LAPIS syntax is defined by an EBNF grammar covering all seven sections, available in the specification document\footnote{\url{https://github.com/cr0hn/LAPIS/blob/main/spec.en.md}}. The grammar is provided for tooling authors; LAPIS is designed to be consumed directly by LLMs without parsing.

\section{Conversion from OpenAPI}
\label{sec:conversion}

The conversion from OpenAPI 3.x to LAPIS follows deterministic rules:

\begin{enumerate}
    \item \textbf{Reference resolution.} All \texttt{\$ref} pointers are resolved before processing.
    \item \textbf{Schema flattening.} \texttt{allOf} schemas are merged by combining properties. For \texttt{oneOf}/\texttt{anyOf}, the most common variant is selected; complex unions are represented as \texttt{any}.
    \item \textbf{Type extraction.} Schemas referenced more than once in \texttt{components/schemas} become named types; schemas referenced once are inlined.
    \item \textbf{Operation mapping.} Each path/method combination becomes an operation. The operation name is derived from \texttt{operationId} or synthesized as \texttt{method\_resource}.
    \item \textbf{Error centralization.} Error responses (4xx/5xx) are collected across all operations, deduplicated by code and schema identity, and emitted once in \texttt{[errors]}.
    \item \textbf{Webhook extraction.} OpenAPI 3.1 \texttt{webhooks} objects are mapped to the \texttt{[webhooks]} section.
    \item \textbf{Metadata discarding.} The following fields are discarded: \texttt{openapi} version, \texttt{info.contact}, \texttt{info.license}, \texttt{info.termsOfService}, \texttt{externalDocs}, \texttt{tags} definitions, \texttt{x-*} extensions, multiple \texttt{servers}, response \texttt{headers}, \texttt{examples}, \texttt{xml} annotations, and \texttt{discriminator}.
\end{enumerate}

A reference converter implementing these rules is available as a Python package (\texttt{pip install lapis-spec}) and as a browser-based tool at \url{https://cr0hn.github.io/LAPIS/}.

\section{Structural Waste in OpenAPI Specifications}
\label{sec:waste}

To motivate the design of LAPIS, we analyzed the structural redundancy in real-world OpenAPI specifications. Table~\ref{tab:errors} shows error response duplication across three APIs.

\begin{table}[H]
\centering
\caption{Error response duplication in OpenAPI specifications.}
\label{tab:errors}
\begin{tabular}{@{}lrrrl@{}}
\toprule
\textbf{API} & \textbf{Ops} & \textbf{Error defs} & \textbf{Unique codes} & \textbf{Most repeated} \\
\midrule
GitHub     & 1,080 & 1,594 & 14 & 404 (531$\times$) \\
Petstore   & 19    & 26    & 3  & 400 (15$\times$) \\
Twilio     & 197   & 1     & 1  & 408 (1$\times$) \\
\bottomrule
\end{tabular}
\end{table}

The GitHub API is particularly illustrative: 14 unique error codes are defined 1,594 times across 1,080 operations. The 404 Not Found response---structurally identical in every occurrence---is repeated 531 times, the 403 Forbidden 349 times, and the 422 Unprocessable Entity 261 times. Each repetition includes the full response schema reference and description.

In LAPIS, these 1,594 definitions are replaced by 14 lines in a single \texttt{[errors]} section. Using the \texttt{cl100k\_base} tokenizer, the GitHub API's LAPIS representation uses 313,101 tokens versus 1,811,843 tokens in OpenAPI YAML---a reduction of 82.7\%.

\section{Evaluation}
\label{sec:evaluation}

\subsection{Methodology}

We evaluated LAPIS against five publicly available API specifications of varying size and complexity. Each specification was obtained from the API provider's official repository or endpoint and converted to LAPIS using the reference converter (\texttt{lapis-spec} v0.1.0, available on PyPI).

Token counts were measured using two tokenizers:
\begin{itemize}
    \item \texttt{cl100k\_base}: Used by GPT-4, GPT-4 Turbo, and GPT-3.5 Turbo (via the \texttt{tiktoken} library).
    \item \texttt{o200k\_base}: Used by GPT-4o and subsequent OpenAI models.
\end{itemize}

For JSON-source specifications, we measured four formats: OpenAPI JSON (formatted), OpenAPI JSON minified, OpenAPI YAML (converted from JSON), and LAPIS. For YAML-source specifications, we measured OpenAPI YAML and LAPIS.

\subsection{Test Corpus}

Table~\ref{tab:corpus} describes the test corpus.

\begin{table}[H]
\centering
\caption{Test corpus: real-world API specifications.}
\label{tab:corpus}
\begin{tabular}{@{}lrrrp{4.5cm}@{}}
\toprule
\textbf{API} & \textbf{Endpoints} & \textbf{Types} & \textbf{Source size} & \textbf{Description} \\
\midrule
GitHub       & 1,080 & 8,313 & 11.8 MB & Full REST API \\
DigitalOcean & 545   & 3,594 & 2.5 MB  & Cloud infrastructure \\
Twilio       & 197   & 1,079 & 1.9 MB  & Communications platform \\
HTTPBin      & 73    & 0     & 28 KB   & HTTP testing service \\
Petstore     & 19    & 33    & 17 KB   & Reference API \\
\bottomrule
\end{tabular}
\end{table}

The corpus spans three orders of magnitude in API size (19 to 1,080 endpoints), covers different industries (cloud infrastructure, communications, developer tools), and includes both JSON and YAML source formats.

\subsection{Token Reduction Results}

Table~\ref{tab:results} presents the primary benchmark results.

\begin{table}[H]
\centering
\caption{Token counts and reduction ratios (\texttt{cl100k\_base} tokenizer).}
\label{tab:results}
\begin{tabular}{@{}lrrrr@{}}
\toprule
\textbf{API} & \textbf{OpenAPI YAML} & \textbf{LAPIS} & \textbf{Reduction} & \textbf{Ratio} \\
\midrule
GitHub       & 1,811,843 & 313,101 & 82.7\% & 0.17$\times$ \\
DigitalOcean & 586,731   & 54,201  & 90.8\% & 0.09$\times$ \\
Twilio       & 306,453   & 24,197  & 92.1\% & 0.08$\times$ \\
HTTPBin      & 6,007     & 1,689   & 71.9\% & 0.28$\times$ \\
Petstore     & 4,634     & 800     & 82.7\% & 0.17$\times$ \\
\midrule
\textbf{Total} & \textbf{2,715,668} & \textbf{393,988} & \textbf{85.5\%} & \textbf{0.15$\times$} \\
\bottomrule
\end{tabular}
\end{table}

The mean token reduction is \textbf{85.5\%} against OpenAPI YAML (the most common human-authored format) and ranges from 71.9\% (HTTPBin) to 92.1\% (Twilio).

\subsection{Cross-Format Comparison}

For JSON-source specifications, Table~\ref{tab:crossformat} shows LAPIS reduction against all OpenAPI variants.

\begin{table}[H]
\centering
\caption{LAPIS token reduction versus OpenAPI variants (\texttt{cl100k\_base}).}
\label{tab:crossformat}
\begin{tabular}{@{}lrrr@{}}
\toprule
\textbf{API} & \textbf{vs JSON} & \textbf{vs YAML} & \textbf{vs JSON-mini} \\
\midrule
GitHub    & 86.6\% & 82.7\% & 79.8\% \\
Twilio    & 93.7\% & 92.1\% & 90.9\% \\
Petstore  & 79.0\% & 82.7\% & 79.0\% \\
\bottomrule
\end{tabular}
\end{table}

LAPIS achieves substantial reductions even against minified JSON, demonstrating that the savings derive primarily from structural redesign rather than whitespace elimination.

\subsection{Tokenizer Consistency}

Table~\ref{tab:tokenizers} compares results across the two tokenizers.

\begin{table}[H]
\centering
\caption{LAPIS token counts across tokenizers.}
\label{tab:tokenizers}
\begin{tabular}{@{}lrrl@{}}
\toprule
\textbf{API} & \textbf{cl100k\_base} & \textbf{o200k\_base} & \textbf{Difference} \\
\midrule
GitHub       & 313,101 & 314,609 & +0.5\% \\
DigitalOcean & 54,201  & 54,463  & +0.5\% \\
Twilio       & 24,197  & 24,324  & +0.5\% \\
HTTPBin      & 1,689   & 1,703   & +0.8\% \\
Petstore     & 800     & 815     & +1.9\% \\
\bottomrule
\end{tabular}
\end{table}

Token counts are highly consistent across tokenizers (within 2\%), confirming that LAPIS's savings are tokenizer-independent and derive from structural rather than encoding-level optimization.

\subsection{Scaling Behavior}

Figure~\ref{fig:scaling} analyzes the relationship between API size and token reduction.

\begin{table}[H]
\centering
\caption{Token reduction versus API complexity.}
\label{fig:scaling}
\begin{tabular}{@{}lrrr@{}}
\toprule
\textbf{API} & \textbf{Endpoints} & \textbf{Types} & \textbf{Reduction} \\
\midrule
HTTPBin      & 73    & 0     & 71.9\% \\
Petstore     & 19    & 33    & 82.7\% \\
GitHub       & 1,080 & 8,313 & 82.7\% \\
DigitalOcean & 545   & 3,594 & 90.8\% \\
Twilio       & 197   & 1,079 & 92.1\% \\
\bottomrule
\end{tabular}
\end{table}

The data shows that token reduction generally increases with API complexity, particularly with the number of types. This is expected: larger APIs have more opportunities for error centralization and type deduplication. HTTPBin, with zero type definitions and minimal error schemas, achieves the lowest reduction because its waste is primarily structural overhead rather than duplication.

\subsection{Character and Line Reduction}

Token reduction is accompanied by proportional reductions in raw character count and line count, confirming that LAPIS is also more compact for storage and human scanning.

\begin{table}[H]
\centering
\caption{Multi-dimensional size comparison.}
\label{tab:dimensions}
\begin{tabular}{@{}lrrrrrr@{}}
\toprule
& \multicolumn{2}{c}{\textbf{Characters}} & \multicolumn{2}{c}{\textbf{Lines}} & \multicolumn{2}{c}{\textbf{Tokens}} \\
\cmidrule(lr){2-3} \cmidrule(lr){4-5} \cmidrule(lr){6-7}
\textbf{API} & \textbf{YAML} & \textbf{LAPIS} & \textbf{YAML} & \textbf{LAPIS} & \textbf{YAML} & \textbf{LAPIS} \\
\midrule
GitHub  & 8.97M & 1.20M & 246K & 18K  & 1.81M & 313K \\
DO      & 2.50M & 200K  & 74K  & 7K   & 587K  & 54K \\
Twilio  & 1.49M & 86K   & 36K  & 3K   & 306K  & 24K \\
HTTPBin & 28K   & 7K    & 1.1K & 258  & 6.0K  & 1.7K \\
Petstore & 22K  & 3K    & 839  & 143  & 4.6K  & 0.8K \\
\bottomrule
\end{tabular}
\end{table}

\section{Discussion}
\label{sec:discussion}

\subsection{Sources of Token Reduction}

The 85.5\% average token reduction derives from four distinct sources, which we estimate based on section-by-section analysis:

\begin{enumerate}
    \item \textbf{Metadata elimination ($\sim$25--30\%).} Discarding fields irrelevant to LLM reasoning: contact information, license, terms of service, tag definitions, vendor extensions, multiple server URLs, examples, and XML annotations.
    \item \textbf{Signature syntax vs. nested structure ($\sim$25\%).} Representing operations as single-line signatures with directional markers (\texttt{>}, \texttt{<}) instead of deeply nested YAML/JSON paths with \texttt{parameters}, \texttt{requestBody}, \texttt{responses}, \texttt{content}, \texttt{application/json}, \texttt{schema} hierarchy.
    \item \textbf{Type system compaction ($\sim$15\%).} Single-line enum definitions, omission of JSON Schema structural keywords (\texttt{type: object}, \texttt{properties:}, \texttt{required:}), and direct type references instead of \texttt{\$ref} paths.
    \item \textbf{Error centralization ($\sim$10--20\%).} Defining each error code once instead of per-operation. The savings scale with API size and error diversity; for GitHub (1,594 error definitions reduced to 14), this is the dominant savings source.
\end{enumerate}

\subsection{Comparison with TOON}

TOON~\cite{toon} and LAPIS operate at different abstraction levels. TOON is a generic data serialization format: it can encode any JSON data more compactly, achieving 30--60\% token reduction through tabular array representation. If TOON were applied to an OpenAPI JSON specification, it would reduce the encoding overhead but the specification would retain all structural redundancies (duplicated errors, nested schemas, irrelevant metadata). LAPIS achieves greater reduction (85.5\%) because it operates at the specification structure level, eliminating domain-specific waste that no generic format can address.

The formats are complementary rather than competing. TOON could encode API response data compactly; LAPIS describes the API itself compactly. They target different content types for different purposes.

\subsection{Information Gain}

LAPIS does not merely subtract information from OpenAPI; it also adds structured representations for concepts that OpenAPI cannot express:

\begin{itemize}
    \item \textbf{Webhook triggers} (\texttt{!} syntax): Semantic conditions that cause webhook delivery, enabling LLMs to predict when events will fire.
    \item \textbf{Structured rate limits}: Machine-readable rate/quota/tier information, enabling LLMs to reason about throttling and retry strategies.
    \item \textbf{Operation flows}: Explicit multi-step sequences showing how operations chain together, enabling LLMs to plan complex API interactions.
\end{itemize}

This means LAPIS provides \emph{more} information per token than OpenAPI: it carries operational semantics that are either absent from or buried in free-text descriptions in the original specification.

\subsection{Limitations}

\paragraph{Lossy by design.} LAPIS deliberately discards information that is useful for non-LLM consumers: detailed JSON Schema validation constraints, response headers, multiple server environments, OAuth flow details, and extensive examples. This makes it unsuitable as a sole specification format.

\paragraph{No LLM comprehension benchmark.} This paper evaluates token reduction quantitatively but does not include a controlled experiment measuring whether LLMs produce better API calls when given LAPIS versus OpenAPI context. Such an evaluation, following the methodology established by TOON's retrieval accuracy benchmarks~\cite{toon}, is planned as future work.

\paragraph{Converter fidelity.} The reference converter handles standard OpenAPI 3.x constructs but may produce suboptimal output for specifications using advanced features such as deeply nested \texttt{oneOf}/\texttt{anyOf} compositions or non-standard extensions.

\section{Cost Implications}
\label{sec:cost}

Token reduction translates directly to monetary savings for applications that include API specifications in LLM context. Table~\ref{tab:cost} estimates per-call costs for including the GitHub API specification as context at representative API pricing (input tokens).

\begin{table}[H]
\centering
\caption{Estimated per-call cost for GitHub API spec as LLM context (input tokens, February 2026 pricing).}
\label{tab:cost}
\begin{tabular}{@{}lrrr@{}}
\toprule
\textbf{Format} & \textbf{Tokens} & \textbf{Cost/call\textsuperscript{*}} & \textbf{Cost/1K calls} \\
\midrule
OpenAPI JSON & 2,330,098 & \$6.99  & \$6,990 \\
OpenAPI YAML & 1,811,843 & \$5.44  & \$5,436 \\
LAPIS        & 313,101   & \$0.94  & \$939 \\
\midrule
\textbf{Savings vs YAML} & & & \textbf{\$4,497/1K calls} \\
\bottomrule
\end{tabular}
\vspace{0.3em}
{\footnotesize \textsuperscript{*}At \$3.00/M input tokens (Claude Sonnet 4.5 pricing).}
\end{table}

For a production application making 1,000 LLM calls per day with the GitHub API in context, LAPIS would reduce context costs by approximately \$4,497 per thousand calls. For smaller APIs, the absolute savings are proportionally lower but the percentage reduction remains consistent.

\section{Conclusion}
\label{sec:conclusion}

We have presented LAPIS, a domain-specific API description format optimized for LLM context consumption. Through empirical evaluation on five production API specifications spanning 19 to 1,080 endpoints, we demonstrated an average token reduction of 85.5\% compared to OpenAPI YAML, with consistent results across tokenizers.

The key insight is that the majority of token waste in API specifications consumed by LLMs is not an encoding problem (addressable by generic compression formats) but a structural problem rooted in OpenAPI's design for a different class of consumer. LAPIS addresses this by redesigning the specification structure itself: centralizing errors, flattening type definitions, using signature-based syntax, and eliminating metadata irrelevant to LLM reasoning.

LAPIS is released as an open specification (CC BY 4.0) with an automated converter from OpenAPI 3.x, a browser-based tool, and a VS Code extension for syntax highlighting. The specification, converter, and all evaluation materials are available at \url{https://github.com/cr0hn/LAPIS}.

\subsection{Future Work}

\begin{enumerate}
    \item \textbf{LLM comprehension evaluation.} Controlled experiments measuring API task accuracy when using LAPIS versus OpenAPI context, following TOON's retrieval accuracy benchmark methodology.
    \item \textbf{Bidirectional conversion.} Investigating partial reconstruction of OpenAPI from LAPIS for round-trip workflows.
    \item \textbf{MCP integration.} Defining a standard workflow for LAPIS as the description layer and MCP as the execution layer.
    \item \textbf{Extended benchmarks.} Evaluation on a larger corpus of production APIs and with additional tokenizers (Anthropic, Google).
\end{enumerate}


\end{document}